\documentclass[useAMS,usenatbib,usegraphicx,letterpaper]{mn2e}
\usepackage{times}
\title[Galaxy Types in the Sloan Digital Sky Survey Using Supervised Artificial Neural Networks]{Galaxy
Types in the Sloan Digital Sky Survey Using Supervised Artificial Neural Networks}

\author[N. M. Ball et al.]{N. M. Ball,$^{1}$\thanks{E-mail: N.M.Ball@sussex.ac.uk}
J. Loveday,$^{1}$ M. Fukugita,$^{2}$ O. Nakamura,$^{2}$ S. Okamura,$^{3}$ J. Brinkmann,$^{4}$ \newauthor
R. J. Brunner$^{5}$
\\ $^{1}$Astronomy Centre, University of Sussex, Falmer, Brighton, BN1 9QJ, UK
\\ $^{2}$Institute for Cosmic Ray Research, University of Tokyo, 5-1-5 Kashiwa, Kashiwa City, Chiba 277-8582, Japan
\\ $^{3}$Department of Astronomy and Research Centre for the Early Universe, School of Science, University of Tokyo,
7-3-1 Hongo, Bunkyo, Tokyo 113-0033, \\ Japan
\\ $^{4}$Apache Point Observatory, P.O. Box 59, Sunspot, NM 88349, USA
\\ $^{5}$Department of Astronomy, University of Illinois, 1002 W Green Street, Urbana, IL 61801, USA}

\begin{document}

\date{Accepted xxxx Received xxxx}

\pagerange{\pageref{firstpage}--\pageref{lastpage}} \pubyear{2003}

\maketitle

\label{firstpage}

\begin{abstract}
Supervised artificial neural networks are used to predict useful properties of galaxies in the Sloan
Digital Sky Survey, in this instance morphological classifications, spectral types and redshifts. By giving
the trained networks unseen data, it is found that correlations between predicted and actual properties are
around 0.9 with rms errors of order ten per cent. Thus, given a representative training set, these
properties may be reliably estimated for galaxies in the survey for which there are no spectra and without
human intervention.
\end{abstract}


\section{Introduction}

The comparison of the observed distribution of galaxies and their properties with that predicted by theory
is an important task in cosmology. In recent years datasets have become available which enable the
comparison to include large samples and detailed galaxy parameters. The Sloan Digital Sky Survey
\citep[SDSS,][]{york2000} provides a dataset of unprecedented size and quality and thus enables significant
improvement in the detail of the comparison.

One can measure an almost limitless number of parameters to describe a galaxy. It is desirable to have as
much information as possible in the fewest parameters, either continuous or discrete. A one parameter galaxy
`type' is particularly convenient. Examples are the well-known Hubble system, or spectral types based on
lines or principal component analysis.

Principal component analysis (PCA), Fisher Matrix and other techniques provide a linear method of reducing the
dimensionality of the parameter space in this way. However galaxy parameters are in general correlated
in non-linear ways, thus a non-linear approach may be more appropriate. Various methods exist, including
non-linear PCA (e.g. {\tt http://www.cis.hut.fi/projects/ica/}), Information Bottleneck \citep{slonim2001},
and artificial neural networks (ANNs). The latter approach is adopted here.

The derived parameters should be physically meaningful, i.e. they should be directly predicted by theories of
galaxy and large scale structure formation, or be related in a quantitative way. For PCA numerous studies
have found that the principal components of galaxy spectra correlate with various physical processes such as
star formation (via absorption and emission line strengths of, for example, the H$\alpha$ line), and to galaxy colour
and morphology. PCA has been applied to the SDSS and yields a one parameter spectral type known as the
eClass \citep{connolly1999}. A similar parameterization, the $\eta$ class, has been made for the 2dF galaxy
redshift survey \citep{madgwick2002}.

Here ANNs in the Matlab Neural Network Toolbox environment ({\tt http://www.mathworks.com/}) are used to map
galaxy parameters from Data Release One (DR1) of the SDSS on to a single continuous `type'. Here we consider
three different types: morphological classification, spectral type and redshift, with standard photometric
parameters as input.

Previous studies involving galaxy classification using ANNs include \citet{storrie1992},
\citet{serraricart1993}, \citet{adams1994}, \citet{lahav1995}, \citet{naim1995}, \citet*{folkes1996},
\citet{lahav1996}, \citet{odewahn1996}, Naim, Ratnatunga \& Griffiths (1997a, b), \citet{molinari1998},
\citet{detheije1999}, \citet{windhorst1999}, \citet{bazell2000}, \citet{bazell2001}, \citet{ball2001},
\citet{goderya2002}, \citet{odewahn2002}, \citet{cohen2003} and \citet{madgwick2003}.
However none of these used a dataset of the size and quality of DR1, or the
Levenberg-Marquardt training algorithm (\S 3), widely used in neural network research.

The layout of the rest of this paper is as follows: in \S 2 the SDSS is summarized and the datasets used are
described. In \S 3 we describe the ANNs; \S 4 presents the results, followed by discussion in \S 5 and
conclusions in \S 6.

\section{Data}

The SDSS is a project to map $\pi$ steradians of the northern galactic cap in five bands
($u$, $g$, $r$, $i$ and $z$) from 3500--8900 \AA.
This will provide photometry for of order $5 \times 10^{7}$ galaxies (\citealt{fukugita1996};
\citealt{gunn1998}; \citealt{lupton2001}; \citealt{hogg2001}; \citealt{smith2002}; \citealt{pier2003}).
A multifibre spectrograph will provide redshifts and spectra for approximately $10^{6}$ of these.
A technical summary of the survey is given in \citet{york2000}.

The data released to the community so far consists of the June 2001 Early Data Release
\citep[EDR,][]{stoughton2002} and the April 2003 Data Release 1 \citep[DR1,][]{abazajian2003}. These
respectively provide photometric parameters and images for one and several million galaxies and spectra for
39,959 and 134,015 galaxies. This paper uses galaxies from DR1.

The SDSS galaxies with spectra consist of a `main', flux-limited sample ($r < 17.77$), with a median
redshift of 0.104 \citep{strauss2002} and a luminous red galaxy sample, approximately volume-limited to $z
\approx 0.4$ \citep{eisenstein2001}. Only the main sample galaxies are used
here.

\subsection{Galaxy Samples}

We used the main galaxy sample from DR1, with sample cuts of reddening corrected $r$-band magnitude $r <
17.77$, confidence in spectroscopic redshift {\tt zConf} $> 0.85$ and spectroscopic object class {\tt
specClass} $=$ {\tt GALAXY} or emission line galaxy {\tt GAL\_EM}. This gave 104,619 galaxies. For each of the
training, test and simulation samples (see \S\ref{sec:anns}) galaxies with severely outlying parameters
($>10\sigma$ from the mean value for the parameter, generally indicative of a measurement error) were
iteratively removed for each parameter in turn. 2,240 were removed in this way, leaving 102,379. See
\S\ref{sec:parameters} for a description of parameters used. Galaxies with outlying target types were
similarly removed. The order in which the parameters are presented may affect the number of galaxies
removed, but the difference is negligible given the small number of objects affected (almost the
same outliers are removed whatever the order of parameter presention). The parameters were individually
normalized to zero mean and unit variance for input into the neural network (see below). For eClass and
redshift the training samples were evened out by binning the galaxies by target type and removing random
galaxies from the most populated bins until the maximum number of galaxies in a bin was twice the mean
number. Bins with less than this were unaffected. This culling ensures that the training of the network is
not dominated by only a small region of parameter space where there are large numbers of galaxies,
which worsens the performance on the rest of the space, and left a total of 98,402 galaxies. The culling
does remove the Bayesian prior of the relative number of each type of galaxy, but the training samples are
large enough that the performance on the test sample is improved rather than hindered. A more sophisticated
method of creating an even sample is to use K-means clustering or a self-organizing map
\citep[e.g.][]{tagliaferri2002}.

\subsection{Galaxy Parameters} \label{sec:parameters}

The parameters used as input to the neural networks, all available in DR1, are shown in Table \ref{table:
params}.

\begin{table*}
\caption{Galaxy parameters used in this paper. Available in the SDSS public Data Release One (DR1), each is
either a direct output of or a simple combination of outputs of the SDSS photometric pipeline.}
\label{table: params}
\begin{tabular}{cl} \hline
Parameter Number 	&Description\\
\hline
1  			&Petrosian radius in $r$ band\\
2  			&50 per cent light radius in $r$ ($R_{50}$)\\
3  			&90 per cent light radius in $r$ ($R_{90}$)\\
4  			&de Vaucouleurs profile radius in $r$\\
5  			&exponential profile radius in $r$\\
6  			&de Vaucouleurs profile axial ratio in $r$\\
7  			&exponential profile axial ratio in $r$\\
8  			&log likelihood of de Vaucouleurs profile\\
9  			&log likelihood of exponential profile\\
10 			&galaxy surface brightness in $r$\\
11 			&concentration index ($R_{50}/R_{90}$) in $r$\\
12--15 			&model $u-g$, $g-r$, $r-i$, $i-z$ colours\\
16--19 			&Petrosian $u-g$, $g-r$, $r-i$, $i-z$ colours\\
20--24 			&model $u$ $g$ $r$ $i$ $z$ magnitudes\\
25--29 			&Petrosian $u$ $g$ $r$ $i$ $z$ magnitudes\\
\hline
\end{tabular}
\end{table*}

The magnitudes are corrected for galactic reddening using the corrections derived from
\citet*{schlegel1998}.

The galaxy images are fitted with the de Vaucouleurs profile \citep{devaucouleurs1948}

\begin{equation} I(r)=I_0~{\rmn{exp}}\{-7.67~[(r/r_{\rmn{e}})^{1/4}-1]\}, \end{equation}
and the exponential profile \citep{freeman1970}

\begin{equation} I(r)=I_0~{\rmn{exp}}(-1.68~r/r_{\rmn{e}}), \end{equation}
where $I_0$ and $I(r)$ are the intensities at radii 0 and $r$, and $r_{\rmn{e}}$ is the half-light radius for the
galaxy. The profiles are truncated to go smoothly to zero at $8r_{\rmn{e}}$ and $4r_{\rmn{e}}$ respectively.

The profile likelihoods are standard $\chi^2$ fits. The model magnitude is that from the better of the two
fits.

The Petrosian magnitude is a modified form of that introduced by \citet{petrosian1976}. It measures a
constant fraction of the total light. The Petrosian flux is given by

\begin{equation} F_{\rmn{P}} \equiv \int_{0}^{N_{\rmn{P}}r_{\rmn{P}}} 2\pi r'dr'I(r') \end{equation}
where $r_{\rmn{P}}$ is the Petrosian radius, which is the value at which the Petrosian ratio of surface brightnesses

\begin{equation} R_{\rmn{P}}(r) \equiv \frac{\int_{0.8r}^{1.25r} 2\pi r'dr'I(r')/[\pi (1.25^2-0.8^2)]}{\int_{0}^{r}2\pi
r'dr'I(r')/(\pi r^2)} \end{equation}
has a certain value, chosen in the SDSS to be 0.2.
The number $N_{\rmn{P}}$ of Petrosian radii within which the flux is measured is equal to 2 in the SDSS.

The magnitude $m$, as with the model magnitude, is then given in asinh units, which are virtually identical
to the usual astronomical magnitudes \citep{pogson1856} at high signal to noise but work at low signal to
noise and negative flux:

\begin{equation} m = - \frac{2.5}{\ln 10} \left [{\rm asinh} \left(\frac{f/f_0}{2b} \right) + \ln b \right], \end{equation}
where $b$ is a softening parameter. Further details are given in \citet*{lupton1999} and \citet{stoughton2002}.

The concentration index is $R_{50}/R_{90}$ where $R_{50}$ and $R_{90}$ are the radii within which 50 and 90 per cent of
the Petrosian flux is received.

The surface brightness used here is given by
\begin{equation} \mu = r + 5\log(\pi r_{\rmn{P}}^2), \end{equation}
$r_{\rmn{P}}$ being the Petrosian radius in the $r$ band.

Parameters other than magnitudes and colours are measured in the $r$ band, since this band is used to define
the aperture through which Petrosian flux is measured for all five bands. Further details of all the
parameters are given on the DR1 webpage ({\tt http://www.sdss.org/dr1}).

\subsection{Target Types}

The networks were separately trained on the following three targets.

\subsubsection{Eyeball Morphological Type}

1875 SDSS galaxies have been classified into morphological types by \citet{nakamura2002}.
The system used was a modified version of the T-type system \citep{devaucouleurs1959}, with the types being
assigned in steps of 0.5 from 0 (early type) to 6 (late type). Unassigned types ($-1$) and galaxies flagged
as being likely to have bad photometry were removed.

The Nakamura et al. catalogue is based on a pre-DR1 version of SDSS data and so their catalogue was matched
to DR1 by equatorial coordinates with a tolerance of 0.36 arcsec, so that the number of duplicate matches is
negligible. This gave 1399 matches.

\subsubsection{eClass}

The eClass is a continuous one parameter type assigned from the projection of the first three principal
components (PCs) of the ensemble of SDSS galaxy spectra. The locus of points forms an approximately one
dimensional curve in the volume of PC1, PC2 and PC3. This is a generalization of the mixing angle $\phi$ in
PC1 and PC2
\begin{equation} \phi = {\rm tan}^{-1} \left( \frac{a_2}{a_1} \right) , \end{equation}
where $a_1$ and $a_2$ are the eigencoefficients of PC1 and PC2.

The range is from approximately $-1$ (corresponding to late type galaxies) to 0.5 (early type).

The eClass is also robust to missing data in the spectra used for its derivation, and is almost independent
of redshift. Further details can be found in \citet{connolly1995}, \citet{connolly1999}, and
\citet{yip2002}.

\subsubsection{Redshift}

The redshift is calculated automatically by the SDSS spectroscopic software pipelines
\citep[][]{stoughton2002}, Frieman et al. (in preparation), and has a success rate of almost 100 per cent.

\section{Artificial Neural Networks} \label{sec:anns}

ANNs, as collections of interconnected neurons each able to carry out simple processing were originally
conceived as being models of the brain. This is still true, however the networks used here are vastly
smaller and simpler and are best described in terms of non-linear extensions of conventional statistical
methods.

The supervised ANN takes parameters as input and maps them on to one or more outputs. A set of vectors of
parameters, each vector representing a galaxy and corresponding to a desired output, or target, is
presented. The network is trained and is then able to assign an output to an unseen parameter vector.

This is achieved by using a training algorithm to minimize a cost function which represents the difference
between the actual and desired output. The cost function $c$ is commonly of the form

\begin{equation} c = \frac{1}{N} \sum_{k=1}^N~(o_{k} - t_{k})^{2},\end{equation} 
where $o_{k}$ and $t_{k}$ are the output and target respectively for the $k$th of $N$ objects.

In general the neurons could be connected in any topology, but a commonly used form is to have an $a: b_1 :
b_2 : \ldots : b_n :c$ arrangement, where $a$ is the number of input parameters, $b_{1 \ldots n}$ are the
number of neurons in each of $n$ one dimensional `hidden' layers and $c$ is the number of neurons in the
final layer, equal to the number of outputs. Here we have one output, $c = 1$. Multiple outputs can give
Bayesian {\it a posteriori} probabilities that the output is of that class given the values of the input
parameters. (This is classification, whereas a single output, $c = 1$, is strictly regression.) Each neuron
is connected to every neuron in adjacent layers but not to any others.

Following \citealt{lahav1996}, each neuron $j$ in layer $s$ receives the $N$ outputs $x^{(s-1)}_{i}$ from
the previous layer $s-1$ and gives a linear weighted sum over the outputs,
\begin{equation} I^{(s)}_j = \sum_{i=0}^{N} w^{(s)}_{ij} x^{(s-1)}_{i}. \end{equation}
There is usually an additive constant, $w_{0j}$, where $x_0 = 1$, in this linear sum. This `bias' allows the
outputs to be shifted in analogy with a DC level.

The neuron then performs a non-linear operation (the transfer function) on the result to give its output
$x^{(s)}_j$, typically a sigmoid or, as used here, the tanh function, which has an output range of $-1$ to
1:
\begin{equation} x^{(s)}_j = \frac{2}{1 + {\rmn{exp}}(-2I^{(s)}_j)} - 1. \end{equation}

The parameters are normalized to zero mean and unit variance. This is not strictly necessary as the net can
in principle perform an arbitrary non-linear mapping, but it enables the weights to be initialized in the
range $-1$ to 1 and not be made unduly large or small relative to each other by the training. This is
particularly helpful for larger networks.

The weights are prevented from growing too large by using weight decay, a regularisation method which adds a
term $d$ to the cost function which penalizes large weights:
\begin{equation} d = \rmn{const} \times \frac{1}{2} \sum_{j} w_{j}^{2}.\end{equation}
Regularisation is also helped by the normalization.

The weights are adjusted by the training algorithm. In galaxy classification this has typically been
the well-known backpropagation algorithm (\citealt{werbos1974}; \citealt{parker1985}; \citealt*{rumelhart1986})
or the quasi-Newton algorithm \citep[e.g.][]{bishop1995}. The Matlab software allows the specification of which
one to use from a number
of choices including these. Here another algorithm popular in neural net research is used: the
Levenberg-Marquardt method (\citealt{levenberg1944}; \citealt{marquardt1963}, also detailed in
\citealt{bishop1995}). This has the advantage that it is very quick to converge to a minimum of the cost
function, and it is able to cope with steep gradients in the parameter-cost function space by approximating
gradient descent, and with shallow gradients by approximating Newton's method. It is thought to be the
fastest algorithm for networks of up to a few hundred weights and its implementation in Matlab further
improves its performance.

Following the neural network toolbox documentation, the algorithm works by using the fact that when the cost
function has the form of a sum of squares the computationally expensive Hessian matrix $\mathbfss H$ can be
approximated as:

\begin{equation} \mathbfss{H} = \mathbfss{J}^{T}\mathbfss{J},\end{equation}
and the gradient is:

\begin{equation} \bmath{g} = \mathbfss{J}^{T}\bmath{e},\end{equation}
where $\mathbfss{J}$ is the (much easier to compute) Jacobian containing the first derivatives of the
network errors with respect to the weights and biases and $\bmath e$ is a vector containing the network
errors, where the network error is the network type minus the target type.

The algorithm then performs the update:

\begin{equation} \bmath{w}_{k+1} = \bmath{w}_{k} - [\mathbfss{J}^{T}\mathbfss{J} + \mu
\mathbfss{I}]^{-1}\mathbfss{J}^{T}\bmath{e},\end{equation}
where $\mathbfss{I}$ is the identity matrix and $\mu$ is the `momentum'. A large $\mu$ approximates gradient
descent and $\mu = 0$ is Newton's method. $\mu$ is given a large initial value so that gradient descent
enables the area of the minimum to be found quickly. It is then decreased after each step where the cost
function reduces, thus moving towards Newton's method which is faster and more accurate near the minimum.

Matlab allows a number of adjustable parameters for the training. The default values were used. The
parameters include:

\begin{description}
\item {\tt epochs}: 	maximum number of training iterations (100)
\item {\tt min\_grad}:	minimum gradient of the cost function ($1.00 \times 10^{-10}$)
\item {\tt mu}:		initial value of $\mu$ ($1.00 \times 10^{-3}$)
\item {\tt mu\_dec}: 	amount to multiply $\mu$ by when the cost function is reduced by a step (0.1)
\item {\tt mu\_inc}: 	similarly for when the cost function increases (10)
\item {\tt mu\_max}:  	maximum $\mu$ value ($1.00 \times 10^{10}$)
\end{description}

The criteria used for stopping training were {\tt epochs}, {\tt min\_grad}, and {\tt mu\_max}, whichever was
reached first. An explanation for {\tt mu\_max} being used is that, whilst appearing indicative of a
diverging solution, it is in fact showing that the algorithm is unable to make a further step to reduce the
cost function. The algorithm only steps if a resulting reduction is found, so it tries progressively larger
steps to search for this, until {\tt mu\_max} is reached. One could also use as a stopping criterion a
validation sample, in which the training is stopped if the cost function when the network at that stage of its
training is run begins to increase. However, with Levenberg-Marquardt the minimum may be reached in very few
iterations (e.g. ten or less), and with the large training samples used here the validation sample gives virtually
the same value of the cost function as the training sample. There is little danger of overfitting because of
the size of the training sample and the intrinsic spread in the galaxy properties. An exception may be a large
network with the eyeball training sample (see \S\ref{sec:archs}).

In general the space of parameters and cost function may have arbitrarily many local minima. It is thus
necessary to start with several random initializations of the weights (or `runs') to avoid a poor local
minimum giving spurious results. The results can then be viewed with the poorest networks down-weighted or ignored,
or by using the
median type. Here the median type is used because although very few runs will be significantly poorer than
average, the ones that are may be by enough such that the mean is a worse measure than the median. The median
type quoted in this paper is always taken from ten runs. The typical scatter between runs is found to be
significantly less than the mean RMS spread of the network types about the targets.

The trained network is then applied to the test sample, and it is for this sample that the tabulated results are
recorded. The training and test samples must be independent but the training sample must be representative of the
test sample. Here the galaxies are given in a random order, the first half was used for training, and the
second half for testing.
For the eClass and redshift one eighth of the DR1 galaxies were used for training and 10,000 for testing.
The samples had their outliers removed, and those with eClass and redshift targets were evened, using the
methods described in \S 2, resulting in training and test samples of approximately 10,000 galaxies (8,501 and
9,801 for eClass; 10,132 and 9,801 for redshift). These samples are easily large enough to train and test the
networks without using undue amounts of memory.
The resulting eyeball samples of 674 (training) and 683 (testing) were not evened as this would make the samples
too small using the method here.
For eClass and redshift the network was simulated on the rest of the DR1 sample with outliers removed
(79,769 galaxies for both targets).

Further details on neural nets can be found in \citet{bishop1995} and in the context of galaxy
classification in \citet{lahav1996}.

\section{Results}

The networks were iterated over many parameter sets, architectures and random initializations of weights.
The results are shown for the parameter sets for the network architectures 1 (single neuron) and 8:1 (8
neurons in a hidden layer) in Table \ref{table: paramscorr}. Some of the best sets (highest correlation between
network output and target type/lowest root mean square difference between network output and target type;
the one almost always corresponds with the other) were run on more architectures. These are shown in Tables
\ref{table: archscorr} and \ref{table: archsrms}. The architectures shown give reasonable execution times,
since the Levenberg-Marquardt algorithm has memory requirements which scale as $N^2$ where $N$ is the number
of weights. The largest number of weights used is in the hundreds.

\begin{table*}
\caption{Correlations and RMSs of median network outputs with target types for the galaxy parameter sets
used in this paper. The first figure is the correlation for a single neuron, the second is for an 8:1
network. Ten runs with random initializations of the weights are used. The RMS is the root mean square
difference between the median network output and the target type. The values are for the neural network
test sample, as opposed to the simulation samples shown in Fig.s \ref{figure: eClass} and \ref{figure: z}
(see \S\ref{sec:anns}, but note that `test' and `simulation' in this context does not mean that the results
are preliminary). The numbers change by amounts of order 0.01 if a different random training sample is used.}
\label{table: paramscorr}
{\scriptsize
\begin{tabular}{lcccccc}
\hline
							&             &Correlation &            &             &RMS         &\\
\hline
Parameter Set 						&Eyeball type &eClass	   &Redshift    &Eyeball type &eClass	   &Redshift\\
Approximate range of targets				&0 to 6	      &-0.5 to 1   &0 to 0.4    &0 to 6	      &-0.5 to 1   &0 to 0.4\\
\hline
Petrosian radius in $r$ band                            &0.492 0.515 &0.096 0.097 &0.266 0.315 &1.291 1.271 &0.195 0.195 &0.050 0.050\\
50 percent light radius in $r$                          &0.567 0.603 &0.162 0.172 &0.312 0.361 &1.221 1.183 &0.193 0.192 &0.050 0.049\\
90 percent light radius in $r$                          &0.296 0.302 &0.044 0.054 &0.212 0.266 &1.416 1.414 &0.196 0.196 &0.051 0.050\\
de Vaucouleurs profile radius in $r$                    &0.802 0.819 &0.366 0.407 &0.423 0.429 &0.886 0.852 &0.180 0.176 &0.047 0.047\\
Exponential profile radius in $r$                       &0.759 0.817 &0.338 0.395 &0.416 0.427 &0.968 0.857 &0.183 0.177 &0.047 0.047\\
de Vaucouleurs profile axial ratio in $r$               &0.493 0.490 &0.084 0.086 &0.292 0.298 &1.290 1.292 &0.195 0.195 &0.050 0.050\\
Exponential profile axial ratio in $r$                  &0.547 0.547 &0.081 0.088 &0.300 0.305 &1.241 1.241 &0.195 0.195 &0.050 0.050\\
log likelihood of de Vaucouleurs profile                &0.051 0.699 &0.212 0.435 &0.381 0.518 &1.481 1.070 &0.191 0.172 &0.048 0.045\\
log likelihood of exponential profile                   &0.131 0.523 &0.230 0.432 &0.222 0.295 &1.471 1.264 &0.190 0.174 &0.051 0.050\\
galaxy surface brightness                               &0.573 0.628 &0.282 0.296 &0.114 0.289 &1.215 1.154 &0.187 0.186 &0.052 0.050\\
concentration index in $r$                        	&0.751 0.782 &0.525 0.534 &0.251 0.281 &0.981 0.927 &0.162 0.161 &0.051 0.050\\
model $u-g$ colour                                      &0.620 0.691 &0.783 0.892 &0.376 0.421 &1.164 1.075 &0.116 0.084 &0.048 0.047\\
model $g-r$ colour                        		&0.492 0.565 &0.804 0.900 &0.711 0.768 &1.309 1.224 &0.113 0.081 &0.037 0.033\\
model $r-i$ colour                        		&0.425 0.558 &0.706 0.739 &0.602 0.636 &1.344 1.231 &0.135 0.128 &0.042 0.040\\
model $i-z$ colour                      		&0.441 0.552 &0.779 0.822 &0.369 0.402 &1.333 1.236 &0.118 0.106 &0.049 0.048\\
Petrosian $u-g$ colour                                  &0.576 0.637 &0.533 0.703 &0.220 0.244 &1.222 1.144 &0.164 0.135 &0.051 0.051\\
Petrosian $g-r$ colour                             	&0.704 0.740 &0.768 0.862 &0.690 0.742 &1.055 0.998 &0.121 0.094 &0.038 0.035\\
Petrosian $r-i$ colour                             	&0.523 0.591 &0.659 0.708 &0.547 0.592 &1.268 1.199 &0.143 0.133 &0.044 0.042\\
Petrosian $i-z$ colour                             	&0.567 0.658 &0.545 0.625 &0.282 0.314 &1.221 1.117 &0.161 0.148 &0.050 0.050\\
model $u$ magnitude 	                                &0.489 0.498 &0.471 0.515 &0.688 0.704 &1.294 1.285 &0.169 0.164 &0.038 0.037\\
model $g$ magnitude                             	&0.243 0.257 &0.155 0.301 &0.644 0.708 &1.438 1.432 &0.193 0.185 &0.040 0.037\\
model $r$ magnitude                             	&0.094 0.155 &0.139 0.147 &0.435 0.437 &1.476 1.464 &0.194 0.194 &0.047 0.047\\
model $i$ magnitude                             	&0.033 0.196 &0.232 0.316 &0.358 0.402 &1.482 1.453 &0.191 0.185 &0.049 0.048\\
model $z$ magnitude                             	&0.042 0.271 &0.332 0.476 &0.298 0.401 &1.481 1.427 &0.184 0.171 &0.050 0.048\\
Petrosian $u$ magnitude                                 &0.529 0.551 &0.436 0.495 &0.628 0.637 &1.259 1.237 &0.173 0.166 &0.041 0.040\\
Petrosian $g$ magnitude                           	&0.310 0.357 &0.189 0.335 &0.662 0.728 &1.410 1.385 &0.191 0.183 &0.039 0.036\\
Petrosian $r$ magnitude                           	&0.111 0.120 &0.102 0.102 &0.467 0.474 &1.473 1.472 &0.195 0.195 &0.046 0.046\\
Petrosian $i$ magnitude                           	&0.040 0.169 &0.196 0.266 &0.391 0.425 &1.481 1.461 &0.192 0.189 &0.048 0.047\\
Petrosian $z$ magnitude                           	&0.080 0.325 &0.307 0.441 &0.325 0.421 &1.478 1.402 &0.186 0.175 &0.049 0.047\\
Petrosian colours $u-g$, $g-r$, $r-i$, and $i-z$	&0.734 0.799 &0.803 0.883 &0.725 0.824 &1.007 0.893 &0.112 0.087 &0.036 0.030\\
Petrosian colours $g-r$ and $r-i$			&0.703 0.759 &0.780 0.863 &0.692 0.761 &1.055 0.966 &0.118 0.094 &0.038 0.034\\
model colours $u-g$, $g-r$, $r-i$, and $i-z$		&0.629 0.753 &0.874 0.936 &0.790 0.881 &1.153 0.978 &0.091 0.065 &0.032 0.025\\
model colours $g-r$ and $r-i$				&0.494 0.620 &0.810 0.904 &0.712 0.789 &1.289 1.163 &0.111 0.080 &0.037 0.032\\
all parameters, except Petrosian and model magnitudes	&0.911 0.928 &0.893 0.943 &0.869 0.922 &0.614 0.554 &0.084 0.062 &0.026 0.020\\
all parameters						&0.911 0.926 &0.893 0.943 &0.870 0.924 &0.615 0.562 &0.084 0.062 &0.026 0.020\\
\hline
\end{tabular}
}
\end{table*}

\begin{table*}
\caption{Correlations for some of the best parameter sets for various ANN architectures using test samples. As in Table
\ref{table: paramscorr}, $\pm$ 0.01 is a representative error on the numbers shown.}
\label{table: archscorr}
{\scriptsize
\begin{tabular}{llcccccccccc}
\hline
	&				&      &      &      &      &\multicolumn{2}{c}{Architecture}\\
\hline
Target	&Parameter Set 			&1     &2:1   &4:1   &8:1   &16:1  &32:1  &4:4:1 &8:8:1 &16:16:1 &8:8:8:1\\
\hline
eyeball&deV and exp radius in $r$	&0.802 &0.820 &0.819 &0.818 &0.817 &0.817 &0.819 &0.819 &0.814 &0.816\\ 
type	&concentration index in $r$	&0.751 &0.781 &0.781 &0.781 &0.784 &0.785 &0.783 &0.785 &0.785 &0.785\\
	&Petrosian $g-r$		&0.704 &0.737 &0.738 &0.739 &0.738 &0.737 &0.738 &0.739 &0.737 &0.738\\
	&Petrosian colours		&0.734 &0.785 &0.790 &0.798 &0.793 &0.762 &0.800 &0.791 &0.744 &0.789\\
	&all except magnitudes		&0.911 &0.920 &0.923 &0.924 &0.920 &0.914 &0.926 &0.925 &0.906 &0.924\\
	&all				&0.911 &0.917 &0.923 &0.920 &0.920 &0.908 &0.922 &0.920 &0.907 &0.914\\

eClass	&model colours			&0.874 &0.931 &0.934 &0.936 &0.936 &0.936 &0.935 &0.936 &0.936 &0.937\\ 
	&Petrosian colours		&0.803 &0.876 &0.879 &0.883 &0.884 &0.884 &0.883 &0.885 &0.884 &0.884\\
	&all except magnitudes		&0.893 &0.935 &0.942 &0.943 &0.944 &0.944 &0.943 &0.944 &0.945 &0.945\\ 
	&all				&0.893 &0.938 &0.942 &0.942 &0.944 &0.944 &0.943 &0.944 &0.945 &0.944\\ 

redshift&model $g-r$			&0.711 &0.759 &0.765 &0.769 &0.769 &0.769 &0.769 &0.769 &0.769 &0.769\\ 
	&model colours			&0.790 &0.860 &0.875 &0.880 &0.885 &0.886 &0.879 &0.886 &0.887 &0.886\\
	&all except magnitudes		&0.869 &0.886 &0.915 &0.923 &0.928 &0.930 &0.918 &0.928 &0.930 &0.929\\
	&all				&0.870 &0.897 &0.915 &0.924 &0.928 &0.930 &0.918 &0.928 &0.929 &0.928\\ 
\hline
\end{tabular}
}
\end{table*}

\begin{table*}
\caption{As Table \ref{table: archscorr}, but showing RMSs.}
\label{table: archsrms}
{\scriptsize
\begin{tabular}{llcccccccccc}
\hline
	&				&      &      &      &      &\multicolumn{2}{c}{Architecture}\\
\hline
Target	&Parameter Set 			&1     &2:1   &4:1   &8:1   &16:1  &32:1  &4:4:1 &8:8:1	&16:16:1 &8:8:8:1\\
\hline
eyeball	&deV and exp radius in $r$	&0.886 &0.849 &0.851 &0.854 &0.856 &0.857 &0.853 &0.853 &0.865 &0.860\\ 
type	&concentration index in $r$	&0.981 &0.929 &0.928 &0.928 &0.923 &0.921 &0.925 &0.921 &0.921 &0.921\\
	&Petrosian $g-r$		&1.055 &1.002 &1.001 &0.999 &1.001 &1.003 &1.001 &1.000 &1.004 &1.000\\
	&Petrosian colours		&1.007 &0.920 &0.910 &0.894 &0.904 &0.965 &0.891 &0.908 &1.008 &0.913\\
	&all except magnitudes		&0.614 &0.582 &0.573 &0.567 &0.581 &0.603 &0.562 &0.565 &0.629 &0.569\\
	&all				&0.615 &0.593 &0.570 &0.581 &0.584 &0.623 &0.576 &0.583 &0.626 &0.604\\

eClass	&model colours			&0.091 &0.068 &0.066 &0.065 &0.065 &0.065 &0.066 &0.065 &0.065 &0.065\\ 
      	&Petrosian colours		&0.112 &0.090 &0.089 &0.087 &0.087 &0.087 &0.088 &0.087 &0.087 &0.087\\
	&all except magnitudes		&0.084 &0.066 &0.062 &0.062 &0.061 &0.061 &0.062 &0.061 &0.061 &0.061\\
	&all				&0.084 &0.065 &0.062 &0.062 &0.061 &0.061 &0.062 &0.061 &0.061 &0.061\\

redshift&model $g-r$			&0.037 &0.034 &0.034 &0.033 &0.033 &0.033 &0.033 &0.033 &0.033 &0.033\\ 
	&model colours			&0.032 &0.027 &0.025 &0.025 &0.024 &0.024 &0.025 &0.024 &0.024 &0.024\\
	&all except magnitudes		&0.026 &0.024 &0.021 &0.020 &0.019 &0.019 &0.021 &0.019 &0.019 &0.019\\
	&all				&0.026 &0.023 &0.021 &0.020 &0.019 &0.019 &0.021 &0.019 &0.019 &0.019\\
\hline
\end{tabular}
}
\end{table*}

\subsection{Effect of Network Architecture} \label{sec:archs}

Tables \ref{table: archscorr} and \ref{table: archsrms} show that a network with a single hidden layer with
a few neurons is adequate for the task of predicting these galaxy parameters using Sloan data. Thus many
network runs could be used to get a good distribution of the assigned type for any particular galaxy. Beyond
about ten hidden neurons there is little improvement and in fact the standard deviation of assigned types to
individual galaxies from the multiple initializations, usually much less than the RMS between actual and
target types, starts to increase. A network, e.g. hidden units of 8:1, is clearly better than a linear
mapping, represented by a single neuron, and although in some cases the improvement in correlation/rms is
not large the plot of network type versus target type (as in Fig.s \ref{figure: eyeball} -- \ref{figure:
z}) is a much smoother function of target type. The networks are almost certainly limited in their
performance by intrinsic scatter in the training sample. This can be seen if the network is tested on the sample
it has just been trained on -- its performance is very similar. This also confirms the earlier statement that
overfitting is unlikely with the 8:1 nets and sizes of training samples used. The increased spread in assigned
types with larger networks may be indicative of overfitting, particularly with the eyeball type as the number
of weights becomes comparable to the number of training examples. The results presented in the figures
use the 8:1 architecture, for which this is not a problem.

\subsection{Effect of Parameter Set}

In general it seems that certain parameters are good for predicting the targets, but that if all the
parameters are added in, the correlation improves over the subsets. The correlation is not improved by
duplicating the few best parameters, so it would appear that genuine information is present in the less good
parameters and it is adding these and not just increasing the size of the network which helps. We therefore use all
parameters in generating the Figures. The model magnitudes used are those from the SDSS DR1 which have been found to
be offset by up to 0.2 mag, but this does not matter here, since the training and test samples are affected in the
same way. As expected, including the magnitudes as well as the colours adds little to the correlation as no
significant new information is added.

\subsection{Results for the Different Target Types}

\subsubsection{Eyeball Morphological Type}

\begin{figure}
\includegraphics[width=8cm]{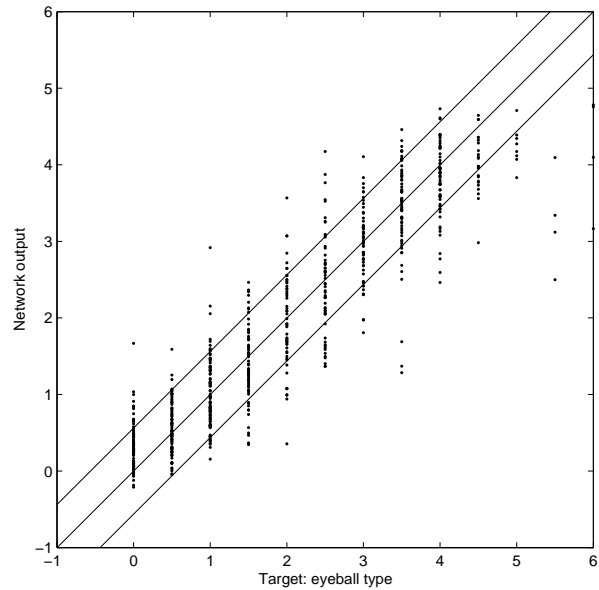}
\caption{Median network type from ten runs versus eyeball morphological type for the eyeball test sample (683
galaxies), using all parameters and the 8:1 network architecture.
The central diagonal line indicates the ideal result, i.e. assigned types equal to the known
type; the diagonal lines above and below are the overall RMS deviation of the network types from the
targets.}
\label{figure: eyeball}
\end{figure}

Previous studies (\citealt{naim1995}; \citealt{lahav1995}) have shown that neural networks are able to
reproduce human-assigned morphological classifications with the same degree of accuracy as another human
expert, about 1.8 types in the $-5$ to 11 T type range. Here the types are assigned in bins of 0.5 in the
range 0 to 6. Fig. \ref{figure: eyeball} shows the median network type versus target type for ten runs.
The network gives correlations up to 0.93 with an RMS of 0.55, about 9 per cent of the range, or the same as the
width of the bins for the types. Smoothing the training sample over the bins by adding random noise of half the
bin width was also tried but this did not improve the correlation, as the bins are quite small relative to
the range in targets.

\subsubsection{eClass Spectral Type}

\begin{figure}
\includegraphics[width=8cm]{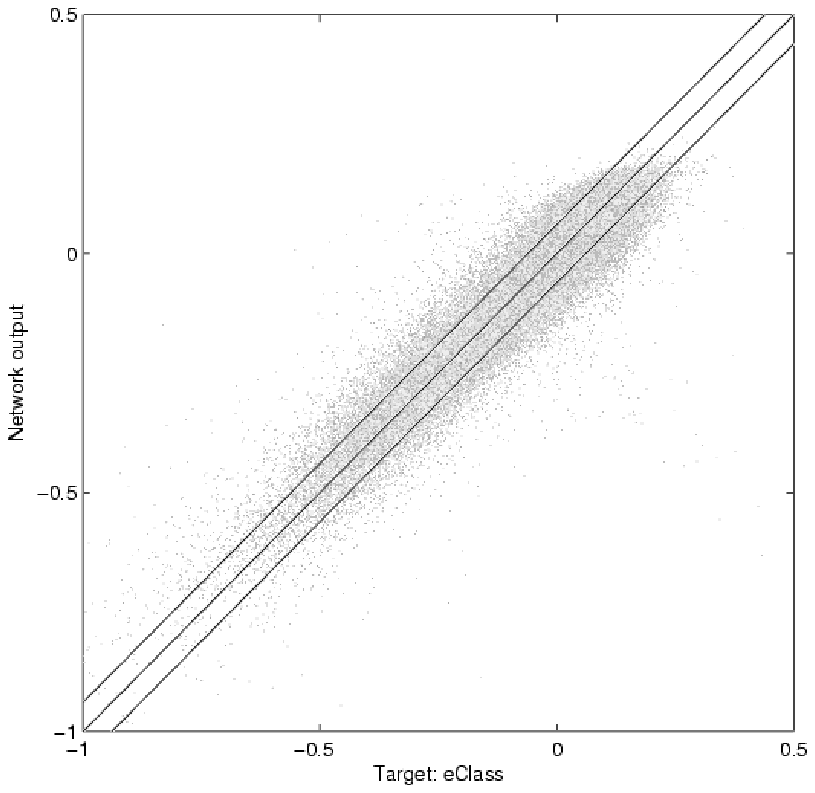}
\caption{Median network type versus SDSS eClass spectral type for the simulation sample of 79,769 DR1 galaxies, using all parameters and the 8:1 network architecture.
The diagonal lines show types equal and $\pm$ the RMS as in Fig. \ref{figure: eyeball}.
Note that the RMS (0.060) and correlation (0.945) are
not identical to those in Table~\ref{table: paramscorr}, as this table shows
results from the smaller test samples.
However the difference is small.}
\label{figure: eClass}
\end{figure}

The ANNs are able to predict the eClass spectral type when trained on galaxies with spectra in the SDSS with
a correlation of up to 0.95 and RMS of 0.06 (4 per cent) for the test sample in the range $-1$ to 0.5. The results
for the simulation on the rest of DR1 are shown in Fig. \ref{figure: eClass}. The shape is not perfect --
a plot of net type - target type versus target type is not precisely symmetrical about zero, but the sigmoid
shape seen when the training sample is not evenly sampled (\S 2) is not as pronounced. The sigmoid shape
has been seen previously, e.g. \citet{naim1995}, where the network `avoided the ends of the scale'.

\subsubsection{Redshift}

\begin{figure}
\includegraphics[width=8cm]{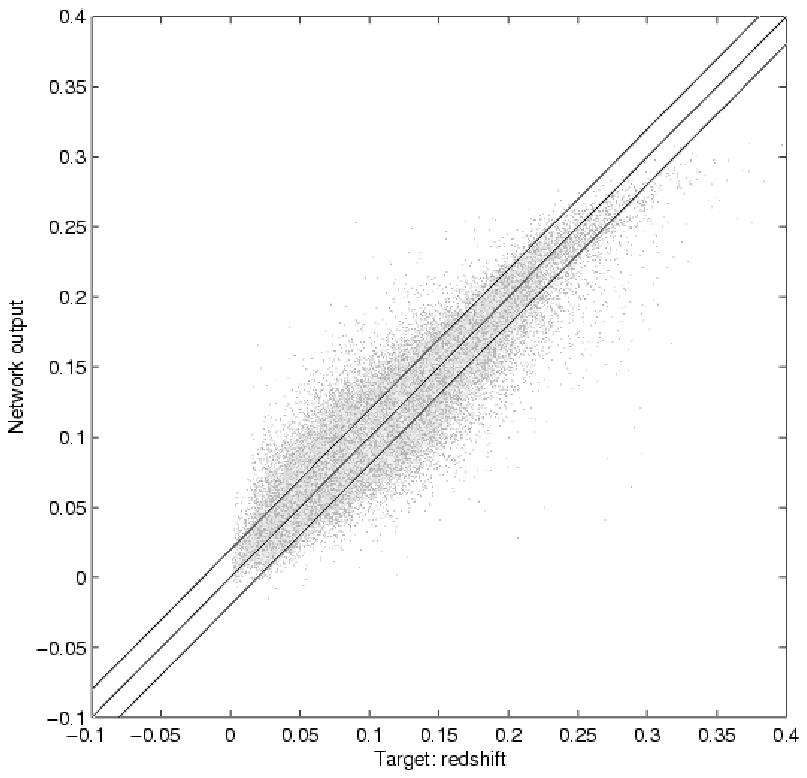}
\caption{As Fig. \ref{figure: eClass}, but with redshift as the target type (again 79,769 galaxies).
For this simulation sample, the RMS is 0.020 and the correlation is 0.925.}
\label{figure: z}
\end{figure}

The generality of the method means that any parameter can be trained on and predicted, hence a photometric
redshift can be obtained (Fig. \ref{figure: z}). The correlation and RMS are up to 0.93 and down to 0.02. The RMS
is comparable to other photometric redshifts in the literature found using neural networks, e.g.
\citet{tagliaferri2002} and \citet*{firth2002}, and to those derived from SDSS data \citep{csabai2003}.

\section{Discussion}

The main result is that the networks can predict morphological classifications, spectral types and
redshifts of galaxies using just photometric parameters. This paper uses moderately sophisticated neural net
techniques on a data set of unprecedented size and quality.
There are many further
techniques which could be used, and possibilities to try out.
In particular there are many sophisticated ANN
techniques which have been little used in astronomy but which may now be justified by the size of the
datasets available. However, with the current data it is unlikely that they would make large improvements as
the results are almost certainly limited by the intrinsic spread in the training samples, and one can never
improve upon the training sample.

Possibilities include, at a basic level, varying the galaxy and neural net parameters used here, for
example the number of random initializations or various Matlab parameters. More sophisticated neural net
techniques include a better even sampling of the training sample so that targets are evenly spread over the range
-- here over-populated bins are simply cut down to size but under-populated bins are not altered. This may
be especially useful for star formation rate and can be done using K-means clustering or a self-organising
map \citep[e.g.][]{tagliaferri2002}. Improved regularisation, e.g. hierarchical Bayesian learning as opposed
to weight decay, could be implemented. Multiple outputs for the network could be used to perform
classification as opposed to regression. This is complementary work rather than an improvement and could be
used for any of the types, in particular the eyeball types, or it could be used to e.g. assign probabilities
to photometric redshift bins, as each output can give the a posteriori probability that the type is that
output given its input parameters. This would also show objects for which the photometric redshift is less
certain, as there may be no one bin with a high probability, or it may be split with peaks occurring in two
separated bins. A different learning algorithm, e.g. quasi-Newton or conjugate gradient, would be needed for
a classifier as Levenberg-Marquardt requires one output. Other learning algorithms could also be used with a
validation sample. The disadvantage of a classifier is that the number of bins for the output is fixed for the
network used. With the regression used here one can bin the assigned types afterwards if desired. Various
methods exist for using committees of networks, apart from that here of using multiple random starts and
using the median type assigned. Examples include constructive learning, bootstrap training samples, forward
selection, backward elimination, cross validation and waterfall. There are other methods for global
optimization apart from multiple random starts, e.g. simulated annealing and genetic algorithms. Many of
these possibilities (regularisation, committees, etc.) are discussed at the comp.ai.neural-nets newsgroup
FAQ at {\tt ftp://ftp.sas.com/pub/neural/index.html}, whilst a waterfall of networks was used in galaxy
classification by \citet{adams1994}. A further possibility is the use of unsupervised networks, i.e. those
in which a predefined similarity criterion is used and the data is left to organize itself, with no training
sample required.  Unsupervised networks objectively find clusters of similar points in a dataset and can be used as a basis
for classification. A well-known unsupervised network which has been used in classifying galaxies is the
Kohonen self-organizing map \citep{kohonen2001}, used by \citet{naim1997b} for galaxy morphology.
\citet{naim1995} used principal component analysis to reduce a set of 24 galaxy parameters to 13 and found
that the latter was as good for predicting types. This was tried, and found to be quicker for networks with
more than ~100 weights over ten runs but it was found that the correlations were generally slightly
worse, as some information was lost (and it cannot be gained by PCA). The time taken was mainly that for the
PCA, then the $N^2$ scaling for Levenberg-Marquardt with $N$ weights.

The usefulness of the methods here is that they are able to predict either spectral parameters using just
photometry or assign morphological types at a vastly greater rate than humans but to the same accuracy. Much
can be done with the types once they have been assigned, and this will form the basis of future work,
in the distributions of these types and in their use to augment large scale structure studies using SDSS
data with other physical measures such as colours.

The SDSS Southern Survey \citep{york2000} is repeatedly imaging a smaller
area of the southern galactic cap to go fainter in imaging and spectroscopy than the northern survey.
Spectra from the Southern Survey could be used as training samples for galaxies at higher redshifts and below the
northern spectroscopic flux limit.

One could thus look at galaxy evolution according to any assigned parameter. One could
also, for example, project galaxies of unknown redshift about ones with known redshift, or push fainter down the
luminosity function if assumptions are made about clustering. A particular statistic of interest is that of
marked point processes \citep*[e.g.][]{beisbart2002}, in which the effects of intrinsic variation and those
of environment can be separated.

Also, if one could predict physical parameters directly this would be extremely useful. One example is the
star formation rate. The sample of 8,683 galaxies detailed in \citet{gomez2002} was investigated. This is a volume
limited sample from $0.05 \le z \le 0.095$ with well measured redshifts and H$\alpha$ star formation rate.
At present the star formation rate is poorly predicted by the ANN, being best at zero but widely spread above this.
Improved results may be obtained for networks trained on just those galaxies which are star forming.

Further targets which could be predicted include the bulge to disc ratio or the 2dF $\eta$ spectral type,
and there are further galaxy parameters which could be used such as S\'ersic indices, or spectral parameters
for predicting morphological types.

It is not immediately obvious whether the resulting distributions say more about the galaxies or the
assigned types, but with the numbers of galaxies available biases in the assigned types from the network
could
be studied in detail. Any biases of this sort are already less than the intrinsic spread in assigned type
and one could compare results using a sample where the target types {\it are} available to see if different
results are obtained. If not, then as long as the sample used has photometry of which the training sample was
representative, the network types can be used with confidence.

\section{Conclusions}
The neural nets are able to predict the eyeball morphological type, the spectral type eClass, and the
redshift using parameters available for all galaxy images in the Sloan Digital Sky Survey Data Release One.
The correlations are 0.93, 0.95, and 0.93 respectively. The mean RMS errors between the network output
and the known type for a set of unseen galaxies of which the training set formed a representative part are
0.55, 0.06 and 0.02 (approximately 9, 4, and 5 per cent of the ranges of the targets).

\section*{Acknowledgments}
Nick Ball thanks Andy Connolly, Andrew Hopkins, Ofer Lahav, Osamu Nakamura,
Bob Nichol, Stephen Odewahn, Andrew Ptak, Kazuhiro Shimasaku, Alex Szalay, Chisato Yamauchi and the comp.ai.neural-nets
and comp.soft-sys.matlab newsgroups for useful discussions and/or information.\\

\noindent Nick Ball is funded by a PPARC studentship.\\

Funding for the creation and distribution of the SDSS Archive has been provided by the Alfred P. Sloan
Foundation, the Participating Institutions, the National Aeronautics and Space Administration, the National
Science Foundation, the U.S. Department of Energy, the Japanese Monbukagakusho, and the Max Planck Society.
The SDSS Web site is http://www.sdss.org/.

The SDSS is managed by the Astrophysical Research Consortium (ARC) for the Participating Institutions. The
Participating Institutions are The University of Chicago, Fermilab, the Institute for Advanced Study, the
Japan Participation Group, The Johns Hopkins University, Los Alamos National Laboratory, the
Max-Planck-Institute for Astronomy (MPIA), the Max-Planck-Institute for Astrophysics (MPA), New Mexico State
University, University of Pittsburgh, Princeton University, the United States Naval Observatory, and the
University of Washington.

\label{lastpage}

\end{document}